\documentclass[sigconf]{acmart}

\input{mysymbol.sty}
\usepackage{amsmath, amsfonts}
\usepackage{algorithmic}
\usepackage{graphicx}
\usepackage{textcomp}
\usepackage[caption=false,font=footnotesize]{subfig}

\usepackage{comment}
\usepackage{multirow}
\usepackage{makecell}
\usepackage{color}
\usepackage{xcolor}
\usepackage{amsthm}
\usepackage{url}
\usepackage{float}
\usepackage{balance}
\usepackage{seqsplit}
\usepackage{xurl}

\AtBeginDocument{%
  }

\setcopyright{acmlicensed}
\copyrightyear{2026}
\acmYear{2026}
\acmDOI{XXXXXXX.XXXXXXX}
\acmConference[E-ENERGY '26]{The ACM International Conference on Future
and Sustainable Energy Systems}{June 23--25,
  2026}{Banff, Canada}
\acmISBN{978-1-4503-XXXX-X/2026/06}

\begin{document}

\title{\texttt{PGLib-CO2}: A Power Grid Library for Real-Time\\Computation and Optimization of Carbon Emissions}


\author{Young-ho Cho}
\email{jacobcho@utexas.edu}
\affiliation{%
  \institution{The University of Texas at Austin}
  \city{Austin}
  \state{Texas}
  \country{USA}
}

\author{Min-Seung Ko}
\email{kms4634500@utexas.edu}
\affiliation{%
  \institution{The University of Texas at Austin}
  \city{Austin}
  \state{Texas}
  \country{USA}
}

\author{Hao Zhu}
\email{haozhu@utexas.edu}
\affiliation{%
  \institution{The University of Texas at Austin}
  \city{Austin}
  \state{Texas}
  \country{USA}
}

\renewcommand{\shortauthors}{cho et al.}

\begin{abstract}
Achieving a sustainable electricity infrastructure requires the explicit integration of carbon emissions into power system modeling and optimization. However, existing open-source test cases for power system research lack generator-level carbon profiling, preventing the benchmark of carbon-aware operational strategies. To address this gap, this work introduces \texttt{PGLib-CO2}, an open-source extension to the \texttt{PGLib-OPF} test case library. The proposed \texttt{PGLib-CO2} enriches standard grid test cases with CO$_2$ and CO$_2$-equivalent emission intensity factors to achieve realistic, generator-level carbon profiling with an expanded list of fuel types.
Using the standardized data, \texttt{PGLib-CO2} allows us to enhance the algorithms for computing key carbon emission metrics, in particular, the locational marginal carbon emissions (LMCE). We first utilize the differentiable programming paradigm for computing LMCE by treating the optimal power flow (OPF)-based grid dispatch as a differentiable layer. This method provides a rigorous marginal sensitivity for general convex cost functions, eliminating the need of using a small incremental change in numerical perturbation. Moreover, to accelerate the real-time LMCE computation, we develop a multiparametric programming (MPP)-based approach that shifts the optimization burden to offline phase of identifying the OPF critical regions. Since each critical region is characterized by a pre-computed affine dispatch function,  the online phase reduces to identifying the region followed by efficiently evaluating the region-specific LMCE values. Numerical evaluations on IEEE test systems demonstrate that the differentiable LMCE computation attains the precise sensitivity information, and the MPP-based approach retrieves the LMCE signals in sub-millisecond timescales—orders of magnitude faster than the direct optimization approach. By bridging high-fidelity data with advanced parametric computation, \texttt{PGLib-CO2} provides a reproducible and computationally efficient foundation for future research in sustainable power system operations.
\end{abstract}

\ccsdesc[500]{Hardware~Power and energy}
\ccsdesc[500]{Hardware~Impact on the environment}

\keywords{Decarbonization, real-time carbon computation, carbon-enriched dataset, differentiable optimization, multiparametric programming}


\maketitle

\section{Introduction}
Electric power grids are among the largest sources of greenhouse gas emissions, serving as crucial infrastructure to power societal and industrial energy demands that are rapidly growing~\cite{poudyal2023resilience}. In particular, the electricity sector produces approximately 40\% of global carbon dioxide (CO$_2$) emissions~\cite{IEA2023}. Environmental and societal concerns have strongly promoted the transition from purely cost and reliability-driven grid operations towards sustainability-aware paradigms that explicitly account for the carbon emission impacts of generating electricity~\cite{olsen2018optimal}. For this reason, \emph{carbon accounting}, which traces electricity back to generation-related emissions, is an increasingly important basis for grid decarbonization planning and market design~\cite{miller2022hourly,chen2024towards}. Hence, accurate carbon modeling for power systems lays the foundation for effective carbon evaluation, informed mitigation strategies, and the design of electricity markets and regulatory mechanisms.

Traditionally, power grid operations have focused mainly on cost minimization and reliability enhancements by employing optimization formulations such as economic dispatch or optimal power flow (OPF) under security constraints~\cite{cain2012history} and linearized nonlinear power flow relations~\cite{cho2023topology,cho2024data}.
Similarly, ongoing efforts on developing synthetic power grid test systems~\cite{Heidel2016GRIDDATA,birchfield2016statistical,birchfield2016grid} or unifying power network data formats~\cite{babaeinejadsarookolaee2019power} mainly focus on traditional metrics such as generation cost functions and network operational limits.
Thanks to a recent effort of the IEEE PES Task Force, a comprehensive standardized set of AC transmission system test cases has been developed, termed as \texttt{PGLib-OPF}, the power grid library for OPF. This library includes IEEE Power Flow test systems and synthetic large-scale test systems \cite{birchfield2016grid}, among other well-known systems, each with complete information on network topology, generation, and load data needed to formulate traditional OPF problems. It not only facilitates the benchmarking across OPF algorithms but also helps to generate large-scale OPF-learning datasets, as in \cite{panciatici2024opfdata,joswig2022opf}.  
However, the \texttt{PGLib-OPF} test systems have not incorporated any carbon emission information for the generation resources, highlighting a critical gap for benchmarking carbon-aware computation and optimization solutions.

Driven by global sustainability concerns, there is a rapidly increasing R\&D effort to incorporate the impact of carbon emissions directly into power system computation and optimization tasks~\cite{kang2012carbon}. 
Taking into account the emission level of CO$_2$, or CO$_2$-equivalent (CO$_2$e) which also includes methane and nitrous oxide, grid operators can better understand and evaluate both economic efficiency and environmental impacts. Recent research~\cite{chen2024carbon,sang2023encoding} has expanded traditional OPF formulations to incorporate carbon emissions costs and constraints, to achieve a cleaner generation portfolio. Meanwhile, exploring load flexibility by shifting flexible demand spatially or temporally could reduce system‐wide emissions~\cite{lindberg2022using,lindberg2021environmental,he2024long}. Despite these promising techniques, the lack of standardized test systems that directly include generator-side carbon information remains a critical barrier for benchmarking various R\&D work.
As a result, almost all research papers have to come up with custom carbon settings or retrofit existing ones using possibly an ad-hoc assignment. 
Recent tools such as \texttt{ElectricityEmissions.jl} can support the calculation of system-wide and location-based carbon emissions, such as the well-known locational marginal carbon emission (LMCE) metric~\cite{gorka2024electricityemissionsjl}. However, they do not provide a systematic approach for generator-level carbon intensity assignments. Therefore, a unified carbon-enriched extension of power system test cases is still missing, yet this is essential for benchmarking the performance of  R\&D work in developing and validating carbon computing and optimization tasks.

Inspired by this need, we develop the \texttt{PGLib-CO2} library, building upon the widely-adopted \texttt{PGLib-OPF} test systems. Our goal is to standardize the settings of CO$_2$ and CO$_2$e \textit{intensity factors} to be consistent with the fuel-type assignments embedded in those test systems. We have further extended the fuel categorization in \texttt{PGLib-OPF} to match established carbon profiling research for general types of generators~\cite{houghton1996climate}. Our open-source tool is packaged for both Python’s \texttt{pandapower} and Julia’s \texttt{PowerModels.jl}, to provide a seamless integration of emission data into grid computation and optimization tasks. This way, it allows users to conveniently adjust the values of the intensity factors and flexibly develop their own use cases or carbon computation tasks.

Inspired by the wide applicability of \texttt{PGLib-CO2} for grid carbon computation, we further develop new algorithms for computing a key carbon emission metric, the locational marginal carbon emissions (LMCE)~\cite{ruiz2010analysis}. This metric is effective for evaluating the potential of emission reduction provided by an incremental load change at any bus location. Thus, it essentially relies on the sensitivity analysis~\cite{li2007dcopf,conejo2005locational} for the optimization problems underpinning the generator dispatch of electricity market such as OPF.  Although the LMCE metric is well-defined, analytically deriving it is limited to specific types of cost functions such as quadratic costs; see e.g.,~\cite{gorka2024electricityemissionsjl}. Accordingly, existing solutions \cite{hawkes2010estimating,he2021using} resort to numerical perturbation by re-solving OPF after a small incremental change. However, the accuracy of this numerical approach could critically depend on the size of incremental change, especially at the boundary operating point between two distinct market congestion patterns~\cite{valenzuela2023dynamic}. To address this issue, we develop a generalized yet accurate LMCE computation method based on differentiable optimization~\cite{griewank2008evaluating,agrawal2019differentiable} by treating OPF as a differentiable layer similar to the standard neural network layer. This way, automatic differentiation provided by the powerful machine learning toolbox can be used to compute the exact emission sensitivities of the OPF outputs with respect to any input change. Hence, our proposed differentiable approach provides a rigorous LMCE computation for general types of cost functions.

Moreover, we pursue a real-time LMCE identification approach with very efficient online computation regardless of the loading condition. Finding the exact LMCE using differentiable programming requires one to solve the OPF problem to construct the computational graph. Under a fast varying grid operating point, this method could be computationally prohibitive and its computation time may be vastly different across the range of operating points. 
To bridge this gap between rigor and real-time usability, we propose a multiparametric programming (MPP)-based approach~\cite{borrelli2003geometric,gal1972multiparametric} that can shift computation offline by identifying the OPF problem's \textit{critical regions}, each with a unique set of active OPF constraints. Accordingly, the online phase reduces to identifying the specific region, followed by efficient mapping to the region-specific LMCE evaluation. This approach can achieve a sub-millisecond identification of locational carbon signals, and possibly allows market participants to indirectly infer the region-based LMCE patterns by using system-wide electricity price signals.

We perform numerical validations of our proposed LMCE algorithms  using the IEEE 14-bus and 118-bus systems. The results confirm that the differentiable method delivers the exact sensitivity values, while the MPP-based approach achieves sub-millisecond retrieval times—orders of magnitude faster than conventional methods—without compromising accuracy.
%

\subsection{Contributions}
In a nutshell, our main contributions are three-fold:
\begin{itemize}
\item \textbf{Standardized Carbon-Enriched Dataset:} \texttt{PGLib-CO2} systematically enriches \texttt{PGLib-OPF} test systems with generator-level carbon attributes (CO$_2$ and CO$_2$e) within \texttt{pandapower} and \texttt{PowerModels.jl} to establish a reproducible foundation for benchmarking carbon-aware grid operations.
\item \textbf{Differentiable LMCE Computation:} We establish a differentiable programming algorithm that computes the exact sensitivity of system-wide emissions to serve as a ground truth for marginal carbon metrics, accommodating general cost functions for which traditional methods may fail.
\item \textbf{Real-Time LMCE Identification via MPP:} We propose a new MPP-based algorithm for LMCE computation that shifts the optimization burden to an offline phase, enabling sub-millisecond retrieval of exact LMCE signals for instantaneous carbon tracking and scalable risk assessment.
\end{itemize}

The rest of the paper is organized as follows. Section \ref{sec:CEM} defines the foundational carbon emission metrics. Section \ref{sec:Dataset} introduces the \texttt{PGLib-CO2} library, detailing the generator-level carbon profiling and the derived carbon-aware optimization frameworks. Section \ref{sec:LMCE} establishes the computational algorithms for LMCE, proposing a differentiable programming framework for exact sensitivity analysis and a multiparametric programming framework for real-time carbon tracking. Section \ref{sec:sim} presents the numerical evaluations using the IEEE 14-bus and 118-bus systems. Finally, Section \ref{sec:con} provides concluding remarks and outlines future research directions.

\begin{figure}[t!]
\begin{center}
\includegraphics[scale=0.5]{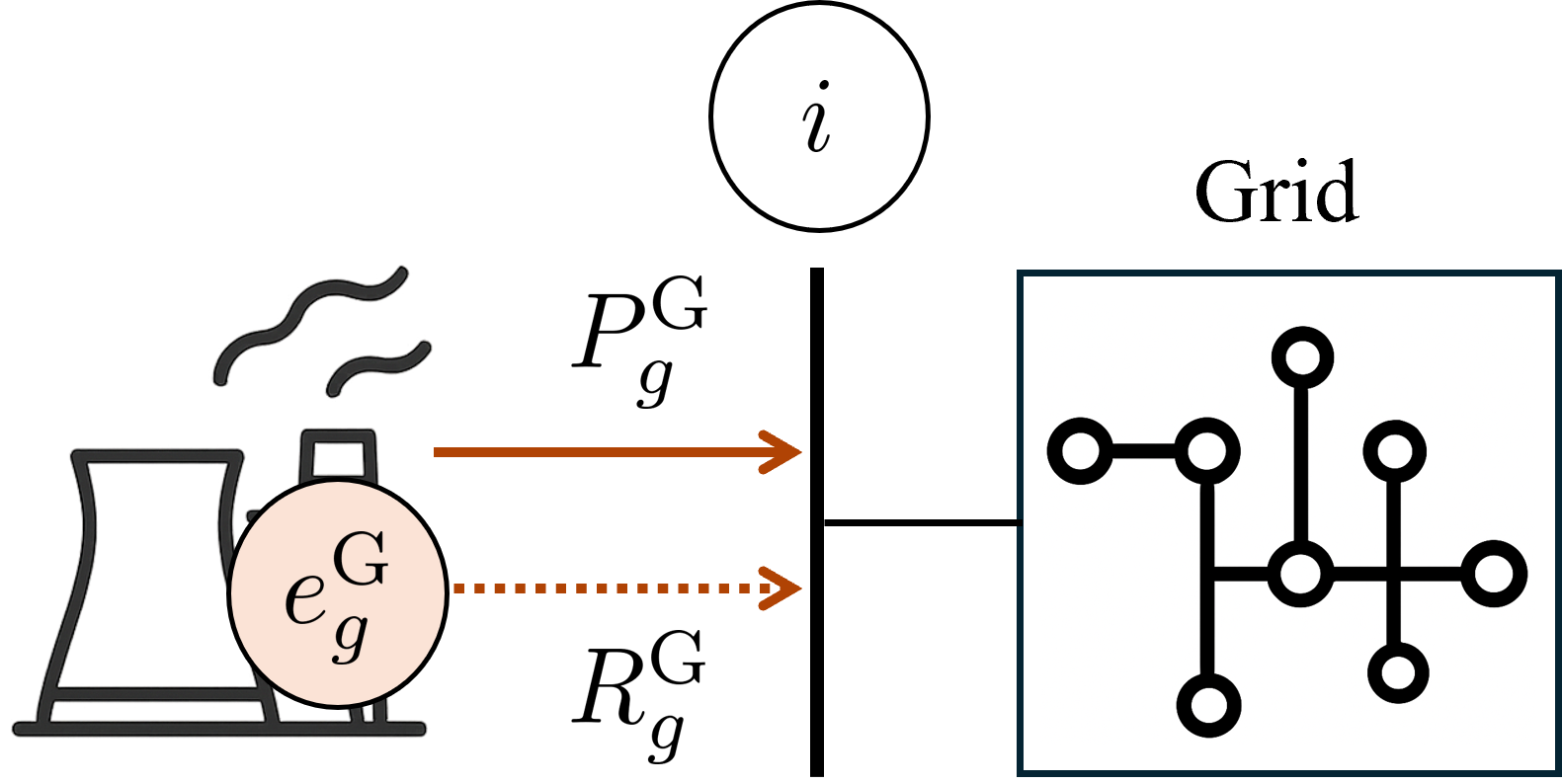}
\caption{Integration of carbon emissions into a power system. Generator $g$ at bus $i$ injects active power, and its carbon emissions can be directly calculated by using the fuel type based emission intensity $e^{\mathrm{G}}_g$.}
\label{structure}
\end{center}
\end{figure}

\section{Carbon Emission Modeling and Metrics} \label{sec:CEM}



Carbon emissions modeling in power systems fundamentally relies on generator-specific emission intensity factors. These factors serve as the critical link between physical power production and environmental impact. As illustrated in Fig.~\ref{structure}, for a generation $g$ connected to bus $i$, we define $e^{\mathrm{G}}_g$ as its constant carbon emission intensity in tons per megawatt-hour (t/MWh). Let $P^{\mathrm{G}}_{g}$ (MW) as the active power output of generator $g\in\ccalG$. The associated carbon emissions are then given by
\begin{equation}
    R^{\mathrm{G}}_{g} = e^{\mathrm{G}}_{g} \cdot P^{\mathrm{G}}_{g},~~\forall g\in\ccalG.
    \label{eq:genEmission}
\end{equation}
When multiple generators are connected to the same bus, their individual $R^{\mathrm{G}}_{g}$ values are aggregated to obtain the total emissions attributed to that bus. This generator-level emission modeling forms the basis for all subsequent emission calculations.

Building on this formulation, we define the \textbf{system-wide total emissions} by summing generator-wise emissions in \eqref{eq:genEmission} across the entire system, as
\begin{align}
    R^{\mathrm{tot}} = \sum_{g \in {\mathcal{G}}} R^{\mathrm{G}}_{g}
                     = \sum_{g \in {\mathcal{G}}} e^{\mathrm{G}}_{g} \cdot P^{\mathrm{G}}_{g}.
    \label{eq:totalEmission}
\end{align}
From this total emission $R^{\mathrm{tot}}$, several widely-used carbon metrics can be computed as listed here. 
\begin{enumerate}
\item \textbf{Average carbon emission (ACE):} Given the total load demand $\sum_{i=1}^N P^{\mathrm{D}}_{i}$, the ACE metric is defined as
\begin{align}
    e^{\mathrm{ACE}}:=\frac{R^{\mathrm{tot}}}{\textstyle \sum_{i=1}^N P^{\mathrm{D}}_{i}}.
\end{align}
ACE captures the average carbon footprint per unit of demand and is commonly used to evaluate system-wide carbon efficiency~\cite{sotos2015amendment}.

\item \textbf{Locational marginal carbon emission (LMCE):} Although the ACE metric captures the system-wide performance, it does not provide the sensitivity of carbon emissions to locational load changes. As an alternative, the LMCE metric can provide this information and is defined with the load at bus $i$, $P^{\mathrm{D}}_{i}$, as
\begin{equation}\label{eq:LMCE}
e^{\mathrm{LMCE}}_i := \frac{dR^{\mathrm{tot}}}{dP^{\mathrm{D}}_{i}}.
\end{equation}
LMCE thus quantifies the sensitivity of total system emissions to marginal load variations in a specific bus~\cite{ruiz2010analysis}, conditioned on the underlying generation redispatch mechanisms.
%

\item \textbf{Locational average carbon emissions (LACE):} To bridge the gap between system-wide averages and marginal sensitivities, several LACE formulations have been proposed to combine both strengths. One notable approach is the network-based \textbf{carbon emission flow (CEF)} model~\cite{kang2012carbon,kang2015carbon}, which allocates carbon emissions by tracing power flows through the network under a proportional power sharing principle.
While this approach is mathematically consistent within its assumptions, it does not explicitly reflect market-driven dispatch decisions and may lead to counterintuitive allocations in meshed networks with loop flows; see \cite{wang2023carbon,chen2024towards} for example system analysis.
To address this issue, more recent dispatch-based approaches integrate marginal emissions along the normalized loading path to compute the LACE metric~\cite{lu2024market}.
Although these methods better align with the market outcomes, their results are highly sensitive to the chosen integration path and may overlook global dispatch effects outside the path. 
Therefore, the definition and computation of LACE still remain open research problems.   
\end{enumerate}


\section{Development of \texttt{PGLib-CO2}}\label{sec:Dataset}
This section presents the development of \texttt{PGLib-CO2} via generator-level carbon profiling, enabling the power system implementations detailed in Section~\ref{sec:usecases}. 
As defined in Section~\ref{sec:CEM}, the carbon metrics computation largely relies on the precise generator-level emission intensity factors. In practice, however, standard power system benchmarks—including the widely adopted \texttt{PGLib-OPF}—typically define generators solely by cost curves and physical limits, without explicit carbon attributes. This lack of standardized emission data creates a gap between carbon modeling methodologies and practical benchmarking, often forcing researchers to rely on ad-hoc or unverified emission assumptions. To address this limitation, \texttt{PGLib-CO2} establishes a reproducible carbon profiling framework that systematically enriches existing test cases with generator-level carbon information.

\subsection{Generator-level Carbon Profiling in \texttt{PGLib-CO2}}
The core of \texttt{PGLib-CO2} lies in associating the active power output of each generator with its carbon emission intensity factors. The intensity factor $e^{\mathrm{G}}_g$ represents the primary carbon characteristic of each generator $g$ and is largely determined by its fuel type~\cite{houghton1996climate}.
For example, clean energy sources such as wind, solar, and hydro have $e^{\mathrm{G}}_{g}=0$, while coal-fired units tend to have the highest $e^{\mathrm{G}}_{g}$ values. Based on this principle, \texttt{PGLib-CO2} expands the limited fuel-type information provided in \texttt{PGLib-OPF}. 
The original dataset includes only a small set of categories, such as distillate fuel oil (\textbf{PEL}), natural gas (\textbf{NG}), bituminous coal (\textbf{COW}), and nuclear (\textbf{NUC}), which is insufficient for detailed carbon profiling~\cite{durvasulu2021data}.
To improve granularity, our \texttt{PGLib-CO2} introduces a richer fuel classification that incorporates additional categories such as anthracite coal (\textbf{ANT}), combined-cycle gas turbines (\textbf{CCGT}), internal combustion engines (\textbf{ICE}), and the aforementioned clean energy sources.

Following research on quantifying carbon emissions, \texttt{PGLib-CO2} also provides two types of intensity factors for each fuel type: CO$_2$ and CO$_2$e emissions. Although CO$_2$ accounts for direct carbon dioxide emissions, CO$_2$e additionally incorporates the impact of methane (CH$_4$) and nitrous oxide (N$_2$O), thus offering a more comprehensive measure of the potential for global warming~\cite{houghton1996climate}. 
Table~\ref{tab:emission_factors} presents the complete list of fuel types and the corresponding intensity factors adopted in \texttt{PGLib-CO2}. These fuel types and factors are derived from well-established research \cite{houghton1996revised}. By providing both CO$_2$ and CO$_2$e metrics, the dataset enables flexible and consistent carbon modeling across a wide range of applications.

\subsection{Workflow for Using \texttt{PGLib-CO2}} 
\texttt{PGLib-CO2} provides a streamlined workflow to enrich the standard \texttt{PGLib-OPF} test cases with generator-level carbon attributes. 
A detailed example is provided in the Appendix.
The workflow begins by loading the original \texttt{PGLib-OPF} network data into one of the two environments, either a Python-based \texttt{pandapower}~\cite{pandapower2018} or a Julia-based \texttt{PowerModels.jl}~\cite{powermodels2018}. Next, one can use the \path{fuel_dict_generation} function to create a dictionary that maps generators to their fuel types and assigns CO$_2$ as the default type of intensity factors.
Note that this dictionary is fully editable, allowing users to modify fuel assignments or switch between CO$_2$ and CO$_2$e.
Finally, this customized dictionary is passed to the \texttt{carbon\_casefile} function, which embeds the fuel type and intensity factor as explicit attributes of the generator database. This step is necessary because fuel-type information in the original \texttt{PGLib-OPF} database is not directly accessible for analysis. 
After conversion, the resulting \texttt{PGLib-CO2} test cases can be directly used for carbon-aware analysis and optimization in both Python or Julia environments.

\begin{table}[t]
\small
\centering
\caption{CO$_2$ and CO$_2$e intensity factors in \textbf{t/MWh} for all fuel types included in \texttt{PGLib-CO2}}
\label{tab:emission_factors}
\begin{tabular}{lcc}
\toprule
Fuel Type\hspace{3cm} & ~~~CO$_2$~~~  & ~~~CO$_2$e~~~  \\
\midrule
Anthracite coal (\textbf{ANT})             & 0.9095  & 0.9143  \\ 
Bituminous coal (\textbf{COW})             & 0.8204  & 0.8230  \\ 
Distillate fuel oil (\textbf{PEL})         & 0.7001  & 0.7018  \\ 
Natural gas (\textbf{NG})                  & 0.5173  & 0.5177  \\ 
Gas combined cycle (\textbf{CCGT})         & 0.3621  & 0.3625  \\ 
Internal combustion engine (\textbf{ICE})  & 0.6030  & 0.6049  \\ 
Nuclear power (\textbf{NUC})               & 0       & 0       \\ 
Renewable energy (wind, solar)             & 0       & 0       \\ 
Hydropower                                 & 0       & 0       \\
\bottomrule
\end{tabular}
\end{table}

\subsection{Use Cases of \texttt{PGLib-CO2}} \label{sec:usecases}

\texttt{PGLib-CO2} supports the direct computation of carbon emission metrics introduced in Section~\ref{sec:CEM}. The library provides built-in functions—such as \path{compute_ACE}, \path{compute_LMCE}, and \path{compute_LACE}—that utilize standardized carbon profiles to explicitly calculate these values. This unified interface allows researchers to perform reproducible environmental diagnostics across various grid test cases, ensuring alignment between theoretical modeling and practical benchmarking.

Beyond metric evaluation, \texttt{PGLib-CO2} also supports carbon-aware optimization by linking generator emission characteristics with economic decision-making. In particular, the emission intensity vector $\bbe^{\mathrm{G}}$ is incorporated into standard optimal power flow (OPF) formulation through a carbon cost function $\bbc(\cdot)$, as defined by
\begin{align}
c (\bbP^{\mathrm{G}}) = (\tau \cdot \bbe^{\mathrm{G}})^\top \bbP^{\mathrm{G}},
\end{align}
where $\tau$ represents a fixed carbon tax rate in dollars per ton of emissions (\$/t). Augmenting the standard OPF objective function with this term enables systematic exploration of the trade-offs between operational efficiency and environmental impact. Note that carbon-related costs can also be considered as constraints and can be imposed at either the system-wide or bus-level, which makes carbon-aware grid optimization more diverse and location-dependent. For detailed discussions, see recent work on carbon-aware OPF~\cite{chen2024carbon,sang2023encoding} and carbon-aware load shifting~\cite{lindberg2021environmental,lindberg2022using,he2024long}.

\section{Algorithms for LMCE Computation} \label{sec:LMCE}
This section presents computational algorithms for determining LMCE. As discussed in Sec.~\ref{sec:CEM}, LMCE requires explicit quantification of the dispatch sensitivity unlike the other metrics.
Consider a transmission system consisting of $N^\mathrm{G}$ generators and $N^{\mathrm D}$ loads, where ${P}^\mathrm{G}_g$ and ${P}^\mathrm{D}_i$ are collected in vectors $\bbP^\mathrm{G} \in \mathbb{R}^{N^\mathrm {G}}$ and $\bbP^\mathrm{D} \in \mathbb{R}^{N^\mathrm {D}}$, respectively.
Using~\eqref{eq:genEmission} and~\eqref{eq:LMCE}, the LMCE vector $\bbe^{\mathrm {LMCE}}$ can be derived by applying the chain rule as
\begin{equation} \label{eq:LMCE_vec}
\bbe^{\mathrm {LMCE}} = \frac{d R^{tot}}{d \bbP^\mathrm{D}} = (\bbe^\mathrm{G})^\top \frac{\partial \bbP^\mathrm{G}}{\partial \bbP^\mathrm{D}} =  (\bbe^\mathrm{G})^\top \bbJ,
\end{equation}
where $\bbJ := \frac{\partial \bbP^\mathrm{G}}{\partial \bbP^\mathrm{D}}$ denotes the dispatch sensitivity matrix. This sensitivity is governed by the underlying market dispatch, formulated here as a convex DCOPF problem:
\begin{subequations}\label{eq:DCOPF}
\begin{align} 
    \min_{\bbP^{\mathrm G}} \quad & f(\bbP^{\mathrm G}) \\
    \text{s.t.} \quad & \mathbf{1}^\top \bbP^{\mathrm G} = \mathbf{1}^\top \bbP^{\mathrm D}, \label{eq:DCOPF_1}\\
    &\underline{\bbF} \leq \bbS(\bbP^{\mathrm G}-\bbP^{\mathrm D}) \leq \bar{\bbF},\label{eq:DCOPF_2}\\
    &\underline{\bbP}^{\mathrm G}\leq \bbP^{\mathrm G} \leq \bar{\bbP}^{\mathrm G},\label{eq:DCOPF_3}
\end{align}
\end{subequations}
where $f(\cdot)$ denotes a generic convex generation cost function (e.g., quadratic or piecewise-linear), $\bbS$ is the injection shift factor (ISF) matrix based on the network topology, $\{\underline{\bbF}, \bar{\bbF}\}$ are the transmission line flow limits, and $\{\underline{\bbP}^{\mathrm G}, \bar{\bbP}^{\mathrm G}\}$ denote the generator power limits, respectively.

Conventional methods for computing $\bbJ$ face critical limitations in both numerical stability and computational speed.
Existing implementations compute LMCE by \emph{numerical perturbation}. They re-solve OPF after applying a small incremental change to the load and approximate $\bbJ$ by finite differences~\cite{hawkes2010estimating,he2021using}. 
While conceptually simple, this approach introduces a step-size trade-off and can become unreliable near boundary operating points where the active congestion pattern changes~\cite{valenzuela2023dynamic}. 
A smaller subset of tools instead derives sensitivities from KKT conditions, which is typically limited to restrictive cost structures (often linear, and only recently extended) and can be numerically fragile under degenerate market conditions~\cite{gorka2024electricityemissionsjl,lu2024market}.

To overcome these barriers, we propose two complementary algorithms. First, we introduce a \textit{differentiable programming} approach (Section~\ref{sec:diff_LMCE}) to compute exact sensitivities for general convex cost structures. Second, to address the fast computation requirements of real-time operations, we propose a \textit{multiparametric programming (MPP)} approach (Section~\ref{sec:mpp}) that enables instantaneous retrieval of LMCE signals.

\subsection{Exact LMCE Computation via Differentiable Programming} \label{sec:diff_LMCE}

\begin{figure}[t!]
\begin{center}
\includegraphics[scale=0.4]{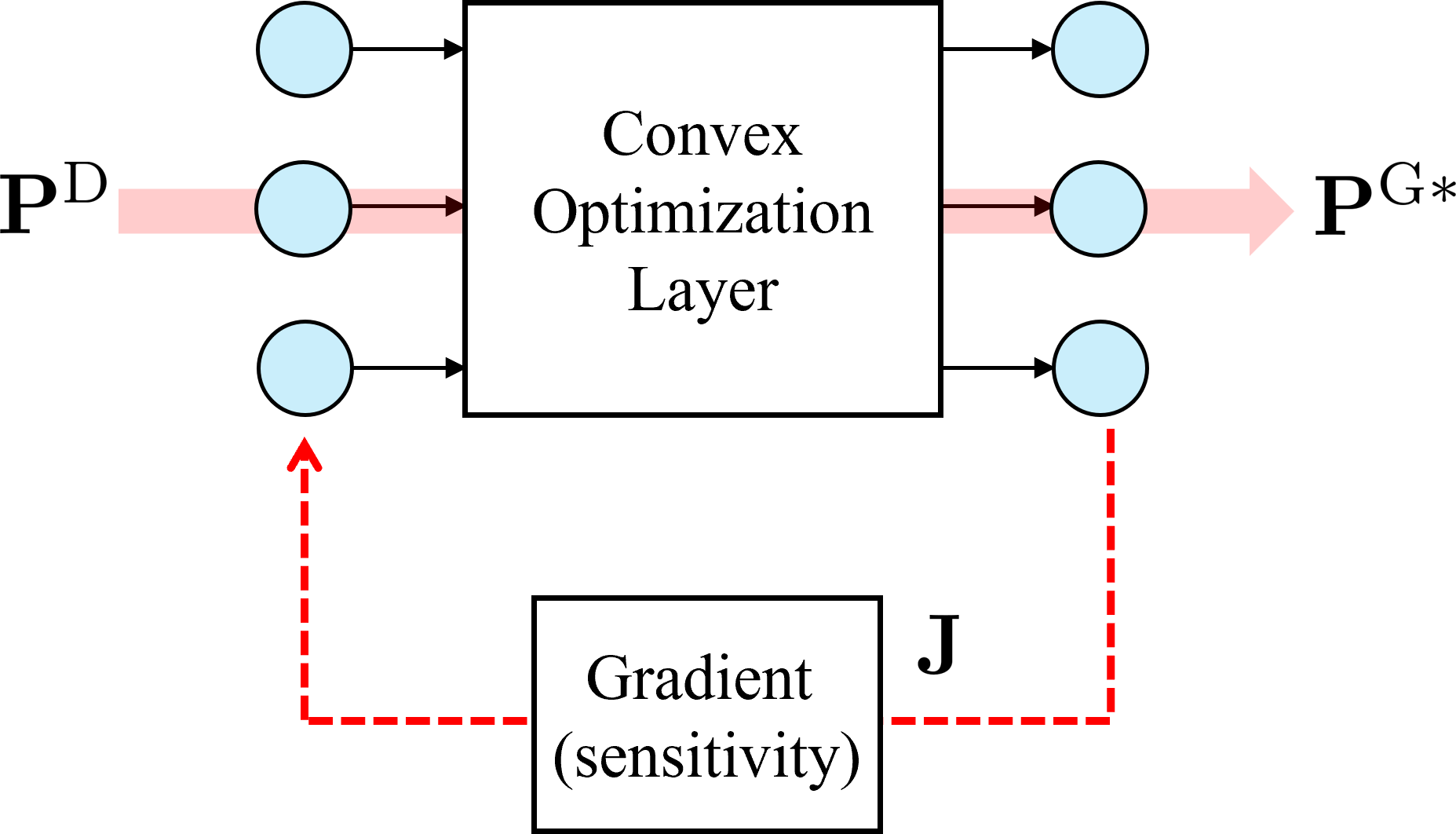}
\caption{Differentiable programming-based sensitivity computation.}
\label{fig:diff}
\end{center}
\end{figure}

We develop a differentiable programming-based computation of $\bbJ$ via \textit{automatic differentiation} (AD)~\cite{griewank2008evaluating}. As illustrated in Fig.~\ref{fig:diff}, we treat the optimization problem as a differentiable optimization layer within a computational graph, following the paradigm of differentiable convex optimizations~\cite{agrawal2019differentiable}. In this formulation, the DCOPF acts as a mapping between input load demands $\bbP^{\mathrm D}$ and output generator dispatches $\bbP^{\mathrm{G}*}$, enabling the Jacobian matrix to be computed automatically via backpropagation engine of deep learning frameworks (e.g., PyTorch).

The core mechanism relies on the implicit function theorem (IFT)~\cite{amos2017optnet,barratt2018differentiability} to propagate gradients through the optimization layer. Unlike standard AD, the derivative of an optimization solution is governed by the set of inequality constraints that are binding at the optimum. Consequently, the sensitivity computation is decoupled into two distinct phases managed by the underlying differential engine:
\begin{itemize}
    \item {\textbf{Forward Pass:}} The solver computes the optimal generator dispatch $\bbP^{\mathrm G*}$ for a given load $\bbP^{\mathrm D}$. Crucially, during this phase, the layer identifies and stores the set of active constraints—the specific inequalities that are satisfied as equalities at the optimum—derived from the primal and dual solutions.
    \item {\textbf{Backward Pass:}} To compute the gradient, the engine does not differentiate the iterative steps of the solver. Instead, it constructs a linear system of equations derived from differentiating the KKT stationarity and complementary slackness conditions restricted to the stored active set. Through the IFT, this construction directly relates perturbations in the parameters ($\bbP^{\mathrm D}$) to changes in the optimal solution ($\bbP^{\mathrm G*}$).
\end{itemize}
Solving this linear system yields the exact Jacobian matrix $\bbJ$. This automated implicit differentiation ensures precise sensitivity information even for general convex cost structures, where the manual derivation of closed-form solutions is often impractical. The LMCE vector is then obtained by projecting the generator emission intensity vector $\bbe^{\mathrm G}$ onto the computed sensitivity $J$, as defined in~\eqref{eq:LMCE_vec}.


\subsection{Real-Time LMCE Identification via Multiparametric Programming} \label{sec:mpp}

Real-time grid operations, particularly instantaneous carbon tracking, require carbon intensity signals that are available on sub-second timescales. Although the differentiable programming in Section~\ref{sec:diff_LMCE} provides the exact $\bbJ$, it remains a locally dependent method that compels the system operator to re-solve the optimization problem for every new load scenario. To eliminate this computational latency, we employ multiparametric programming (MPP)~\cite{borrelli2003geometric,gal1972multiparametric} to shift the optimization burden entirely to an \emph{offline} pre-computation phase. By explicitly characterizing the optimal generator dispatch as a piecewise-affine function of the system load over a predefined load domain $\mathcal{D}$, MPP reduces the \emph{online} task to simple region identification and affine evaluation. In this work, we focus on the multiparametric linear programming (MPLP) formulation derived from a linear-objective DCOPF. Extensions to quadratic objectives (MPQP) follow similar principles but are left for future study.

\begin{itemize}
\item{\textbf{Offline Multiparametric Formulation:}}
We begin by reformulating the DCOPF constraints into the standard multiparametric canonical form that separates the decision variables $\bbP^{\mathrm G}$ from the uncertain parameters $\bbP^{\mathrm D}$. This approach has been widely used for real-time prediction of locational marginal prices (LMPs)~\cite{ji2015probabilistic,ji2016probabilistic}.
In particular, the physical constraints are rewritten such that the constraint matrix $\bbA$ and vector $\bbb$ are invariant, while the nodal demand enters solely through the affine term $\bbU\bbP^{\mathrm D}$, as the form of 
\begin{equation}
    \bbA \bbP^{\mathrm {G}} \le \bbU \bbP^{\mathrm {D}} + \bbb
    \label{eq:canonical}
\end{equation}
By expanding the power balance, transmission flow, and generation capacity constraints, we derive the specific block-matrix structure for this canonical form:
\begin{equation}
    \underbrace{
    \left[ \begin{matrix} 
    \mathbf{1}^\top \\ -\mathbf{1}^\top \\ \bbS \\ -\bbS \\ \mathbf{I} \\ -\mathbf{I} 
    \end{matrix} \right]}_{\bbA} \bbP^{\mathrm {G}}
    \le  
    \underbrace{
    \left[ \begin{matrix} 
    \mathbf{1}^\top \\ -\mathbf{1}^\top \\ \bbS \\ -\bbS \\ \mathbf{0} \\ \mathbf{0} 
    \end{matrix} \right]}_{\bbU} \bbP^{\mathrm {D}}
    +
    \underbrace{
    \left[ \begin{matrix} 
    0 \\ 0 \\ \bar{\bbF} \\ -\underline{\bbF} \\ \bar{\bbP}^{\mathrm G} \\ -\underline{\bbP}^{\mathrm G} 
    \end{matrix} \right]}_{\bbb} .
\end{equation}


A key implication of this canonical structure is that the optimal dispatch is fully determined by the set of active constraints, denoted by $\ccalA$. The KKT stationarity and complementary slackness conditions enforce these constraints as linear equalities:
\begin{align}
    \bbA_{\mathcal{A}}\bbP^{\mathrm {G}} = \bbU_{\mathcal{A}}\bbP^{\mathrm {D}} + \bbb_{\mathcal{A}}
\end{align}
where the subscript $\mathcal{A}$ selects rows corresponding to the active constraints. 
Once the active set $\mathcal{A}$ is identified, the original optimization problem reduces to a determined linear system. This allows us to replace the iterative re-evaluation of solvers with a direct analytical derivation. For any valid active set $\mathcal{A}$, the optimal dispatch $\bbP^{\mathrm {G}*}$ admits an explicit affine representation of the load:
\begin{equation} \label{eq:sensitivity}
    \bbP^{\mathrm {G}*} = \bbJ \bbP^{\mathrm {D}} + \bbg,
\end{equation}
where the sensitivity matrix defined in \eqref{eq:LMCE_vec} becomes $\bbJ = \bbA_{\mathcal{A}}^{-1}\bbU_{\mathcal{A}}$ and the intercept equals to $\bbg = \bbA_{\mathcal{A}}^{-1}\bbb_{\mathcal{A}}$. Both quantities can be calculated offline by inverting a non-singular basis of the active constraints. The existence of such a basis can always be ensured; rigorous treatments of primal and dual degeneracy are provided in~\cite{borrelli2003geometric}.
This confirms that, with the same active set, the generator response follows a deterministic linear policy regardless of the load demand $\bbP^D$.

However, since the set of binding constraints inevitably changes as the load $\bbP^{\mathrm{D}}$ varies, this linear policy is valid only within a bounded domain. We define this domain as the critical region $\mathcal{CR}_k$, which corresponds to the subset of the parameter space $\mathcal{D}$ where the given active set $\mathcal{A}_k$ remains optimal. Geometrically, this region is constructed as a closed convex polytope, as shown in Fig.~\ref{fig:MPP}(a), defined by the intersection of primal and dual feasibility conditions. Primal feasibility requires that the affine dispatch solution defined by the region's specific parameters ($\bbJ_k, \bbg_k$) must not violate any inactive constraints $j \notin \mathcal{A}_k$. Substituting the solution into the original inequalities in~\eqref{eq:canonical} yields the hyperplane boundaries $(\bbA_j \bbJ_k - \bbU_j)\bbP^{\mathrm {D}} \le \bbb_j - \bbA_j \bbg_k$. Dual feasibility further requires the Lagrange multipliers associated with the active constraints to remain non-negative; however, since the cost vector in DCOPF is constant, this condition is typically satisfied throughout the entire polyhedron.

With these properties established, the offline phase constructs the full solution map via a graph-based exploration strategy. The process begins by identifying the optimal active set for a nominal base case. We select a representative load vector $\tilde{\bbP}^{\mathrm{D}} \in \mathcal{D}$ (e.g., the nominal demand) and solve the standard DCOPF to obtain an initial active set $\mathcal{A}_0$. Using~\eqref{eq:sensitivity}, the corresponding sensitivity matrix $\bbJ_0$ and intercept $\bbg_0$ are computed, yielding the "seed" region $\mathcal{CR}_0$. We then initialize a queue with the facets of this seed region.
For each unexplored facet, we apply a basis pivoting rule to define the adjacent region's active set $\mathcal{A}_{new}$, which reflects the transition of a constraint from inactive to binding. If the new active set has not been encountered before, we compute its optimal dispatch function and explicitly store its defining polyhedral boundaries as the inequality set $\mathbf{M}_k \mathbf{P}^{\mathrm{D}} \le \mathbf{k}_k$ (see Fig.~\ref{fig:MPP}(b)). We then add its unique facets to the queue, and the procedure is repeated until all facets either connect to known regions or coincide with the global parameter boundaries. The outcome is a complete lookup table consisting of dispatch policies $\{\bbJ_k, \bbg_k\}$ and region definitions $\{\bbM_k, \bbk_k\}$ ready for online retrieval.

\begin{figure}[t!]
	\centering
	\subfloat[]{\includegraphics[scale=0.25]{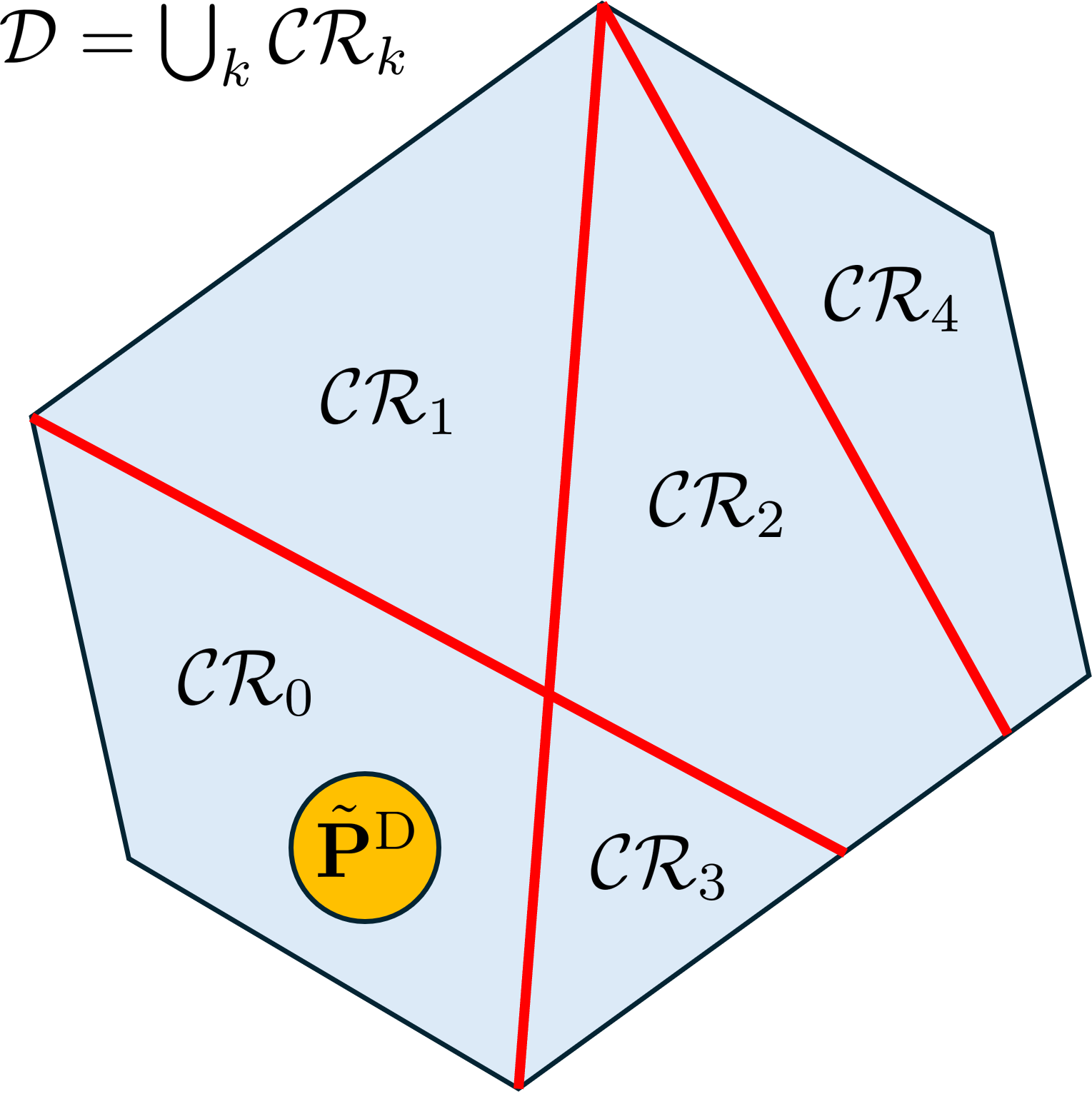}}\quad
    \subfloat[]{\includegraphics[scale=0.25]{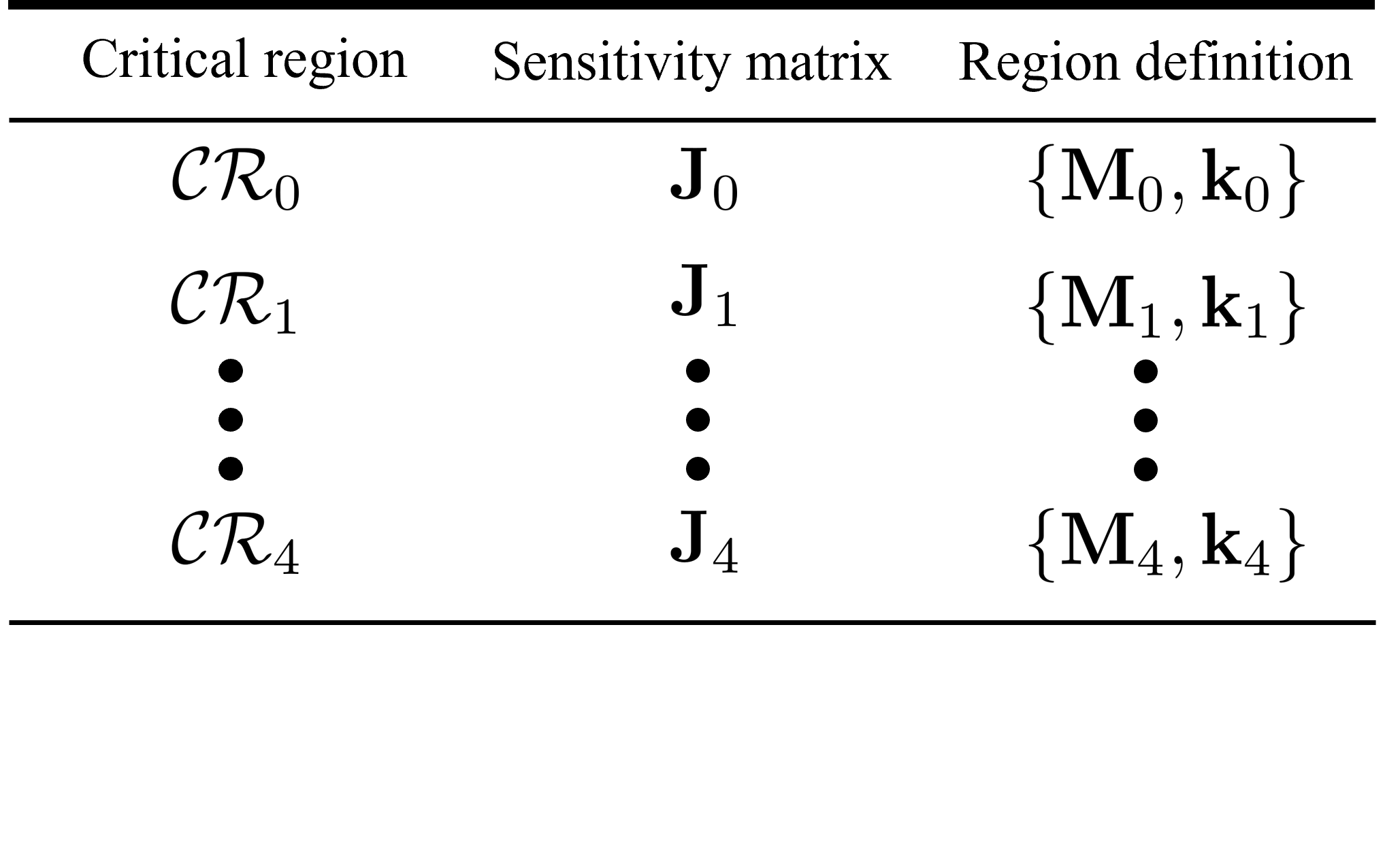}}
	\caption[]{\small (a) The load domain $\mathcal{D}$ is partitioned into disjoint critical regions. (b) The resulting lookup table maps each region $\mathcal{CR}_k$ to its specific sensitivity $\bbJ_k$ and polyhedral boundaries $\{\bbM_k, \bbk_k\}$.}\label{fig:MPP}
\end{figure}

\item{\textbf{Online LMCE Retrieval:}} In the online phase, the pre-computed solution map converts the optimization problem into a geometric point-location problem~\cite{bemporad2002explicit}. For any realized real-time load vector $\hat{\bbP}^{\mathrm D}$, the task reduces to identifying the specific critical region index $k$ such that $\hat{\bbP}^{\mathrm D} \in \mathcal{CR}_{k}$. Since the regions form a disjoint partition, the active region is uniquely determined. We identify the active region $k$ by verifying if the current load satisfies the polyhedral definition $\bbM_{k} \hat{\bbP}^{\mathrm D} \le \bbk_{k}$, where $\bbM_k$ and $\bbk_k$ represent the stored boundary hyperplanes in Fig.~\ref{fig:MPP}(b).

Once the active region is identified, we can retrieve the LMCE vector immediately. Since the dispatch sensitivity $\bbJ_k$ is constant within the critical region, the marginal response of total emissions $R^{\mathrm{tot}}$ to load variations is pre-determined. The exact LMCE vector $\bbe^{\mathrm{LMCE}}$ is therefore obtained by projecting $\bbe^{\mathrm G}$ onto the retrieved sensitivity matrix:
\begin{equation}\label{eq:Online_LMCE}
    \bbe^{\mathrm{LMCE}} = (\bbe^{\mathrm G})^\top \bbJ_{k}
\end{equation}
This retrieval mechanism yields orders-of-magnitude speedups compared to conventional sensitivity analysis. While standard sensitivity analysis requires solving a full linear program with cubic complexity $\mathcal{O}(N^3)$ for every new scenario, the MPP lookup method involves only matrix-vector multiplications with complexity linear to the number of region-defining constraints, $\mathcal{O}(N)$. This efficiency enables high-frequency carbon tracking and probabilistic risk assessment, facilitating real-time decision-making in carbon-aware grid operations.

Beyond enabling high-speed retrieval, the MPP-based method fundamentally bridges the gap between economic signals and environmental metrics. This structure allows the pre-computed lookup table to serve as a decentralized carbon tracking tool. Within a critical region, the optimal cost is affine, $\pi^* = \bbf^\top (\bbJ_k \bbP^{\mathrm{D}} + \bbg_k)$, implying that the LMP vector $\bblambda$ is simply the projection of the generator cost vector $\bbf$ onto the dispatch sensitivity matrix:
\begin{equation}\label{eq:Online_LMP}
\bblambda^\top = \frac{\partial \pi^*}{\partial \bbP^{\mathrm{D}}} = \bbf^\top \bbJ_{k}.
\end{equation}
Recall that this gradient vector is unique to the active constraint set, it can effectively index the lookup table. Consequently, individual loads—who typically lack visibility into system-wide operating conditions—can map posted LMP signals immediately to the pre-computed $\bbJ_k$. This allows market participants to retrieve the exact LMCE based solely on public price data, without requiring access to the operator's internal dispatch information.

\end{itemize}

\section{Numerical Results}\label{sec:sim}

We demonstrate the capabilities and generalizability of \texttt{PGLib-CO2} through the numerical evaluations of LMCE computation tasks using the IEEE 14-bus and 118-bus test systems. We utilize the network data from \path{pglib_opf_case14_ieee} and \path{pglib_opf_case118_ieee} provided in \texttt{PGLib-OPF}. To better demonstrate the carbon impact, we have slightly modified the generator fuel types provided in \texttt{PGLib-OPF}, as the original fuel types tend to have very similar emission intensities and thus lead to limited spatial variability. Specifically, we reconfigure each system by assigning generators to one of three fuel types: high-emission anthracite coal (ANT), mid-emission natural gas (NG), or low-emission gas combined cycle (CCGT). This modification increases the variability of the emission profile within each system, thereby making it more critical to have accurate LMCE values. The corresponding CO$_2$e emission factors for these three fuel types are given in Table~\ref{tab:emission_factors}.

For the 14-bus system, we have introduced some network modifications to induce more frequent congestion patterns, as the original parameters rarely exhibit active transmission constraints. We restricted the thermal limits of the $L=20$ transmission lines in MW to \{70, 90, 50, 70, 50, 20, 50, 70, 90, 90, 20, 70, 50, 70, 20, 50, 90, 50, 50, 70\}, following the configuration in~\cite{kekatos2014grid}. In addition, we have linearized the generator cost functions and adjust their upper output limits. Table~\ref{tab:14_bus} shows these parameters and the specific fuel-type assignments. For the 118-bus system, we have retained the original network settings but updated the generator fuel types according to the specific assignments listed in Table~\ref{tab:118_bus}. All numerical tests have been performed on a laptop with an Intel\textsuperscript{\textregistered} CPU @ 2.70 GHz and 32 GB of RAM.

\begin{table}[t!]
\small
\centering
\caption{Generator parameters and fuel type assignments for the IEEE 14-bus system}
\label{tab:14_bus}
\begin{tabular}{cccc}
\toprule
Generator Bus & \makecell[c]{Linear Cost\\ Coefficient [\$/MWh]}  & Fuel Type & \makecell[c]{Upper Limit\\(MW)} \\
\midrule
1      & 18   & ANT                & 200\\ 
2      & 31   & NG                 & 140\\ 
3      & 30   & NG               & 100\\
6      & 15   & CCGT   & 100\\
8      & 22   & CCGT   & 100\\
\bottomrule
\end{tabular}
\end{table} 

\begin{table}[t!]
\small
\centering
\caption{Generator fuel type assignments for the IEEE 118-bus system}
\label{tab:118_bus}
\begin{tabular}{cc}
\toprule
Fuel Type  & Generator Bus \\
\midrule
ANT    & \makecell[c]{\{10, 26, 46, 49, 59, 61, 80, 89, 100\}}\\ 
NG     & \{25, 31, 54, 69, 103\}\\ 
CCGT   & \{12, 65, 66, 87, 111\}\\
\bottomrule
\end{tabular}
\end{table}

\begin{figure*}[t!]
	\centering
	\subfloat[Load 1]{\includegraphics[scale=0.3]{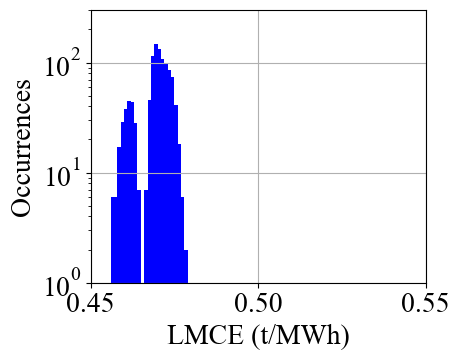}}\quad
    \subfloat[Load 2]{\includegraphics[scale=0.3]{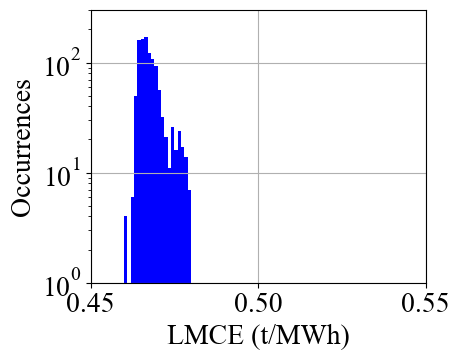}}\quad
    \subfloat[Load 3]{\includegraphics[scale=0.3]{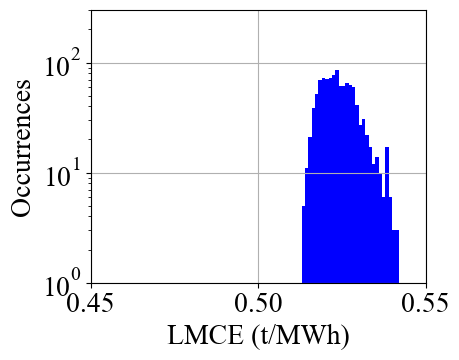}}\quad
    \subfloat[Load 4]{\includegraphics[scale=0.3]{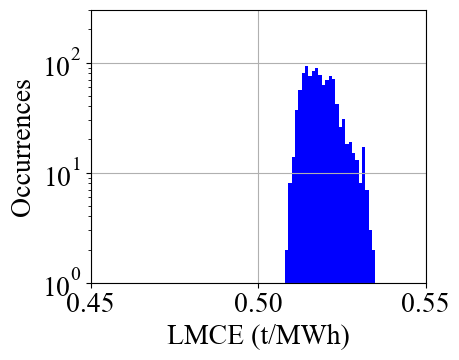}}\\
    \subfloat[Load 5]{\includegraphics[scale=0.3]{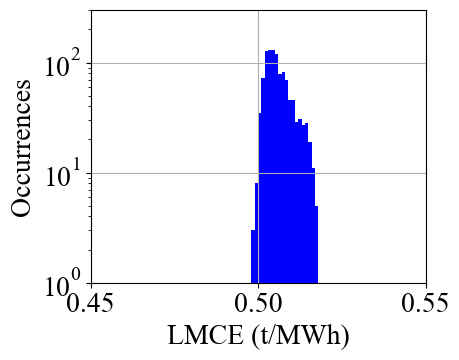}}\quad
    \subfloat[Load 6]{\includegraphics[scale=0.3]{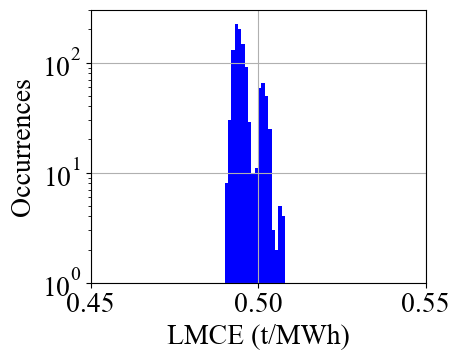}}\quad
    \subfloat[Load 7]{\includegraphics[scale=0.3]{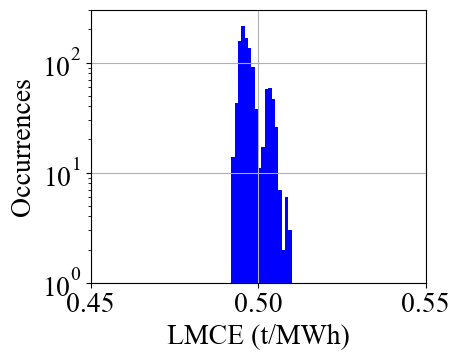}}\quad
    \subfloat[Load 8]{\includegraphics[scale=0.3]{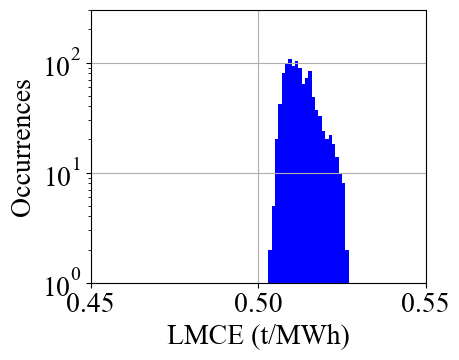}}
	\caption[]{\small Distribution of the LMCE under quadratic generation cost functions for the IEEE 14-bus system}\label{fig:LMCE_quad}
\end{figure*}

\subsection{Validations of LMCE Calculation} \label{sec:result_calc}
We first evaluate the accuracy and versatility of the proposed method on differentiable LMCE computation as established in Section~\ref{sec:diff_LMCE}. The method has been implemented using the \texttt{CVXPYLayer} library to construct the differentiable DCOPF optimization layer, embedded within \texttt{PyTorch} to enable automatic gradient computation through forward and backward passes.
To validate the fidelity of our gradient-based LMCE, we compare our results against the existing \path{ElectricityEmissions.jl} library~\cite{gorka2024electricityemissionsjl}, which was developed for linear generation cost functions only. We generate testing datasets comprising 1,000 distinct load scenarios for both test systems. For the IEEE 14-bus system, we uniformly perturb the demand of the 8 non-generator load buses within the range of $[80, 120]\%$ of their nominal values. For the IEEE 118-bus system, we apply the same perturbation range to the 8 buses with the highest loading levels to induce a sufficient level of grid congestion.


Table~\ref{tab:LMCE_Calc} summarizes the deviations between the LMCE values computed by our proposed differentiable method and those obtained from \path{ElectricityEmissions.jl}. The results demonstrate that two methods achieve nearly identical accuracy, with discrepancies remaining at an operationally negligible level. Specifically, the average deviation remains below 0.021 across both cases. Given that the maximum generator emission factor is 0.9143 (ANT), this corresponds to an average distortion of less than 2.2\%, confirming that the proposed differentiable framework accurately preserves marginal sensitivity information. 
One notable observation is that the maximum deviation for the IEEE 14-bus case reaches 1.648. This outlier occurs exclusively at the boundary operating points between two different critical regions, where the change of active constraint set introduces non-smooth transitions in the dispatch solution. Such boundary points are well known to cause inconsistent sensitivity calculation for manual derivations or numerical perturbations. In contrast, our framework is guaranteed to compute the exact sensitivity at these boundary points via automatic differentiation. The observed discrepancy therefore is caused by the lack of precision in the existing library,  underscoring the need to have accurate and robust LMCE calculation as offered by our proposed method.

\begin{table}[t!]
\small
\centering
\caption{Deviation of LMCE calculation compared to \texttt{ElectricityEmissions.jl}.}
\label{tab:LMCE_Calc}
\begin{tabular}{ccc}
\toprule
  & 14-bus & 118-bus\\
\midrule
Mean deviation  & 0.021  & 0.015 \\ 
Max deviation   & 1.648  & 0.085 \\ 
\bottomrule
\end{tabular}
\end{table}

Beyond the standard linear generator cost functions, we further demonstrate the capability to accommodate general cost structures. In particular, quadratic cost functions could arise  in certain practical scenarios but are not widely supported by existing libraries. We evaluate the LMCE under the quadratic cost functions defined in the original MATPOWER case files for the IEEE 14-bus system. Fig.~\ref{fig:LMCE_quad} illustrates the resulting LMCE distributions across the load scenarios. Unlike the region-based, discrete sensitivity characteristic of linear DCOPF, quadratic costs lead to a continuously varying dispatch pattern~\cite{Chenetal2022}.
Although a direct numerical comparison is infeasible as existing libraries do not support this setting, we have observed that the computed emission factors in Fig.~\ref{fig:LMCE_quad} remain consistently bounded within the maximum and minimum of generator intensities, namely, the range of [0.3625,0.9143] based on Table~\ref{tab:emission_factors}. This consistency supports the validity of  our proposed method when used for quadratic generator cost functions. 


\subsection{MPP-based LMCE Identification}

We further evaluate the effectiveness and computational efficiency of the MPP-based LMCE identification approach established in Section~\ref{sec:mpp}. The algorithm is implemented using the MATLAB-based Multi-Parametric Toolbox (MPT3)~\cite{herceg2013multi} to formulate the multiparametric DCOPF problem and compute the explicit critical regions. The estimated LMCE values are compared with the actual LMCE values obtained by the differentiable method. This comparison utilizes the same testing datasets used in Section~\ref{sec:result_calc}.

\begin{figure}[t!]
\begin{center}
\includegraphics[scale=0.32]{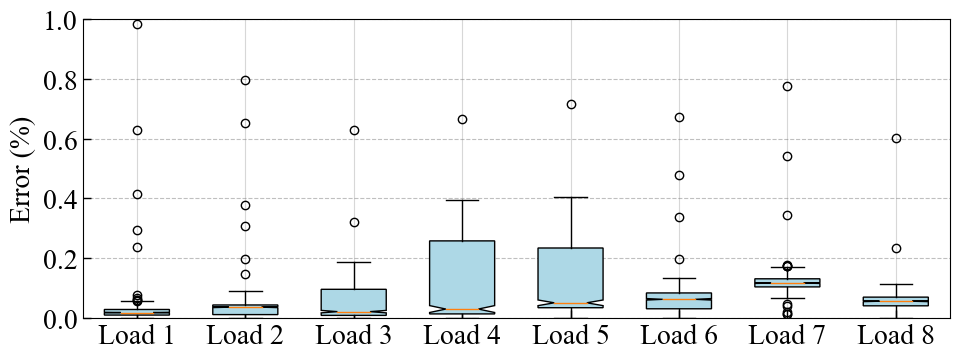}
\caption{Distribution of LMCE estimation errors on the IEEE 14-bus system using the proposed MPP framework.}
\label{fig:LMCE_est}
\end{center}
\end{figure}

We first consider the IEEE 14-bus system. The MPP-based framework solves the problem in~\eqref{eq:canonical} for each loading scenario which was obtained by perturbing the nominal load within the range of $[80, 120]\%$. This step has identified a total of 15 critical regions. Fig.~\ref{fig:LMCE_est} presents the box plots of the estimation error percentages, as normalized by the nominal load, obtained by the online identification stage across all perturbed load scenarios at each load bus. Each box spans the first (Q1) and third (Q3) quartiles, along with the median midline. Data points beyond 150\% of the interquartile range (Q1--Q3) are marked as outliers. We have observed that the errors for all load scenarios largely remain below 0.2\% and the maximum outliers do not exceed 1\%, confirming the extreme high accuracy in identifying the actual LMCE values.

The primary advantage of this identification approach lies in its superior and consistent computational efficiency, as shown by the computation time statistics in Table~\ref{tab:LMCE_Est}. While the differentiable LMCE computation on average requires a computation time of 37.7~ms per scenario, the MPP-based online identification reduces this to merely 0.17~ms. Interestingly, the MPP framework exhibits negligible difference between the mean and maximum computation times, while the maximum time for the differentiable method almost triples the average time. The consistent computation complexity of the MPP-based approach across multiple runs stems from the deterministic nature of its online phase, which reduces to a geometric point identification problem. By simply checking linear inequalities to identify the critical region, the MPP-based approach performs a predictable number of operations regardless of the varying load scenarios. In contrast, the differentiable method relies on iterative optimization solvers where convergence time depends on the numerical conditioning of each scenario. As a result, its maximum computation time spikes to 81.2~ms from the mean of 37.7~ms. This fluctuation reflects the inherent latency risks associated with solving differentiable programming problems from scratch.

\begin{table}[t!]
\small
\centering
\caption{Computation time comparison between Differentiable and MPP frameworks}
\label{tab:LMCE_Est}
\begin{tabular}{ccc|cc}
\toprule
& \multicolumn{2}{c|}{14-bus} & \multicolumn{2}{c}{118-bus} \\ \midrule
& Differentiable & MPP &  Differentiable & MPP \\ \midrule
Mean  & 37.7 ms  & 0.17 ms   & 7.23  s   & 0.14 ms \\ 
Max   & 81.2 ms  & 0.21 ms   & 25.12 s   & 0.16 ms \\ 
\bottomrule
\end{tabular}
\end{table} 

\begin{figure}[t!]
\begin{center}
\includegraphics[scale=0.32]{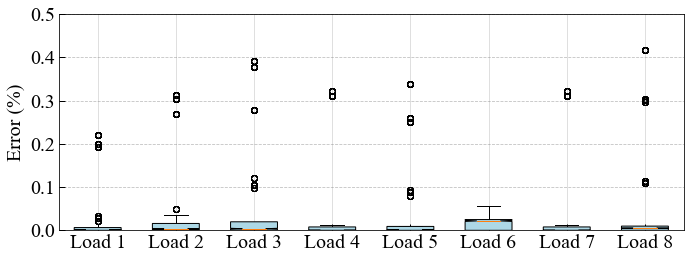}
\caption{Distribution of LMCE estimation errors on the IEEE 118-bus system using the proposed MPP framework.}\label{fig:LMCE_est_118}
\end{center}
\end{figure}

This advantage becomes even more pronounced in the larger IEEE 118-bus system and confirms the excellent scalability of the MPP-based approach. Using the same $[80, 120]\%$ perturbation range, the MPP solver has identified 10 critical regions. Fig.~\ref{fig:LMCE_est_118} shows the box plots of estimation error percentages where the errors remain consistently below 0.1\%. Notably, the computational improvement is more evident, as shown in Table~\ref{tab:LMCE_Est}. The differentiable method requires a total of 130 minutes to process the full dataset of 1,000 load scenarios, which could be computationally prohibitive. Its mean computation time rises to 7.23~s, with a maximum of 25.12~s for the most difficult scenario. In contrast, the MPP-based online phase maintains the millisecond-level computation with a mean of 0.14~ms and a maximum of 0.16~ms. Interestingly, the identification time is even lower than that of the 14-bus system. This reduction is largely driven by the decreasing number of critical regions from 15 to 10, and thus it is not illogical.  This result confirms that the complexity of the online phase depends primarily on the depth of the region search tree rather than the system size or the DCOPF problem dimensionality. This fast and consistent computation time offered by the proposed MPP-based identification approach makes it particularly attractive for real-time applications where worst-case latency is a critical constraint.

\section{Conclusions and Future Work}\label{sec:con}


We developed \texttt{PGLib-CO2}, an open-source extension of the widely used \texttt{PGLib-OPF} library that systematically augments standard power system test cases with generator-level carbon attributes. By bridging existing data gaps, this work establishes a reproducible data foundation for benchmarking carbon-aware operational strategies and unified emission metrics.
Building on these standardized data, we focused on improving the computation of LMCE. We first formulated a differentiable DC OPF layer that enables exact locational marginal carbon emissions (LMCE) computation via implicit differentiation, supporting general convex cost functions beyond linear settings.
%
Furthermore, to meet strict real-time requirements, we proposed a multiparametric programming (MPP) framework for instantaneous LMCE retrieval. This framework shifts the optimization burden offline by partitioning the demand domain into critical regions and strong region-wise dispatch maps. Therefore, this structure enables instantaneous online retrieval of LMCE via region lookup and matrix evaluation without accessing private operator dispatch information.
Numerical evaluations on IEEE systems confirm that the differentiable approach ensures mathematical exactness and the MPP approach achieves orders-of-magnitude speedups. These results demonstrate that \texttt{PGLib-CO2} can serve as a practical benchmark suite not only for carbon accounting, but also for real-time carbon-aware control and optimization.

Several directions merit further investigation. First, extending the MPP framework to quadratic objectives (MPQP) would broaden the applicability to market models with a nonlinear convex structure. Future releases of \texttt{PGLib-CO2} can also incorporate richer generator metadata and scenario sets that reflect renewable variability, storage participation, and extreme operational conditions. 


\begin{acks}
This work has been supported by NSF Grants 2130706 and 2150571.
\end{acks}

\bibliographystyle{ACM-Reference-Format}
\bibliography{ref}

\appendix
\section{Illustrative Code Snippets for \texttt{PGLib-CO2} Utilization}
\label{sec:appendixA}
This appendix provides illustrative code snippets in Python (\path{pandapower}) and Julia (\path{PowerModels.jl}) to demonstrate a streamlined workflow for using \texttt{PGLib-CO2}.
The Python example uses the IEEE 14-bus system, while the Julia example uses the IEEE 118-bus system.
First, the \texttt{fuel\_dict\_generation} function creates a dictionary that maps the generators to their fuel types, assigning CO$_2$ as the default intensity factor.
Then, the examples modify the fuel type of generator 1 (Python) and generator 6 (Julia) to natural gas (NG), updating their intensity factors from the default CO$_2$ to CO$_2$e.
This customized dictionary is passed to the \texttt{carbon\_casefile} function, which embeds the fuel type and intensity factor as formal attributes within the generator data structure.
Finally, a carbon-neutral OPF is solved using the functions provided by \texttt{pandapower} and \texttt{PowerModels.jl}, and the OPF solution along with the corresponding total carbon emission $R^\mathrm{tot}$ is printed.
Through the carbon-enriched extension provided by \texttt{PGLib-CO2}, these resultant test cases allow seamless integration of standard test networks (such as the IEEE 14-bus or 118-bus systems) into fully carbon-aware datasets, enabling researchers to compare baseline and carbon-aware dispatch outcomes.

\subsection*{Python Implementation}

\begin{table}[H]
\centering
\begin{tabular}{l}
\toprule
\small
\makecell[l]{\textbf{Example:} Enriching generator carbon characteristics\\
             \qquad\qquad~ and OPF execution using \texttt{pandapower}}\\
\midrule
\begin{minipage}[t]{0.96\linewidth}\scriptsize
\begin{verbatim}
import pandapower as pp
import pandapower.networks as pn

# 1) Load the IEEE 14-bus test network
net = pn.case14()

# 2) Generate a custom dictionary for
#    generator fuel types and intensity factors
fuel_dict = fuel_dict_generation(net)

#    Manually update fuel type and intensity factor
fuel_dict[1] = {"type": "NG", "emissions": "CO2e"}

# 3) Embed the fuel type and intensity factor
#    as formal attributes within the generator data
carbon_casefile(net, fuel_dict)

# 4) Solve carbon-neutral OPF and analyze results
pp.runopp(net)

# 5) Analyze final dispatch and emissions
print(net.res_gen)
\end{verbatim}
\end{minipage}\\  \bottomrule
\end{tabular}
\end{table}

\subsection*{Julia Implementation}
\begin{table}[H]
\centering
\begin{tabular}{l}
\toprule \small
\makecell[l]{\textbf{Example:} Enriching generator carbon characteristics\\
             \qquad\qquad~ and OPF execution using \texttt{PowerModels.jl}}\\
\midrule
\begin{minipage}[t]{0.96\linewidth}\scriptsize
\begin{verbatim}
using PowerModels
using Ipopt

# 1) Load the IEEE 118-bus test network
net = PowerModels.parse_file("case118.m")

# 2) Generate a custom dictionary for
#    generator fuel types and intensity factors
fuel_dict = fuel_dict_generation(net)

#    Manually update fuel type and intensity factor
fuel_dict[6] = Dict("type" => "NG",
                    "emissions" => "CO2e")

# 3) Embed the fuel type and intensity factor
#    as formal attributes within the generator data
carbon_casefile(net, fuel_dict)

# 4) Solve carbon-neutral OPF and analyze results
result = run_opf(net, ACPPowerModel,
                Ipopt.Optimizer)

# 5) Analyze final dispatch and emissions
println(result["solution"]["objective"])
\end{verbatim}
\end{minipage}\\
\bottomrule
\end{tabular}
\end{table}

\section{Computing Carbon Emission Metrics with \texttt{PGLib-CO2}}
\label{sec:appendixB}

\texttt{PGLib-CO2} provides \texttt{compute\_ACE}, \texttt{compute\_LMCE}, and \texttt{compute\_LACE} to evaluate the carbon metrics defined in Section~\ref{sec:CEM}.
In addition to returning a scalar system-wide summary (ACE), the toolkit supports two computational backends for each locational metric:
\texttt{compute\_LMCE} can be computed via \texttt{"diff"} or \texttt{"MPP"}, and
\texttt{compute\_LACE} can be computed via \texttt{"CEF"} or \texttt{"Riemann"}.

\paragraph{How to choose the backend.}
Select \texttt{"diff"} for \texttt{LMCE} when an exact sensitivity at a specific operating point is needed and per-query optimization/differentiation is acceptable.
Select \texttt{"MPP"} when repeated real-time queries over a fixed demand domain are required; it shifts computation offline and returns LMCE by region lookup and affine evaluation.
For \texttt{LACE}, select \texttt{"CEF"} when a fast average-emissions signal derived from a carbon-emission-factor construction is necessary, and select \texttt{"Riemann"} when you approximate dispatch-based LACE by integrating marginal emissions along a chosen demand scaling path.

\subsection*{Python: ACE/LMCE/LACE Computation}

\begin{table}[H]
\centering
\begin{tabular}{l}
\toprule
\small
\makecell[l]{\textbf{Example:} Compute carbon metrics using \texttt{pandapower}}\\
\midrule
\begin{minipage}[t]{0.96\linewidth}\scriptsize
\begin{verbatim}
import pandapower as pp
import pandapower.networks as pn

# --- Build carbon-enriched case (Appendix A) ---
net = pn.case14()
fuel_dict = fuel_dict_generation(net)
carbon_casefile(net, fuel_dict)

# --- Compute carbon metrics ---
R_tot = compute_Rtot(net)                 # total emissions
ace   = compute_ACE(net)                  # system-wide average (ACE)

# Locational metrics with selectable backends:
lmce_diff = compute_LMCE(net, method="diff")
lmce_mpp  = compute_LMCE(net, method="MPP")

lace_cef     = compute_LACE(net, method="CEF")
lace_riemann = compute_LACE(net, method="Riemann")

print("R_tot =", R_tot)
print("ACE   =", ace)
print("LMCE(diff) =", lmce_diff)
print("LMCE(MPP)  =", lmce_mpp)
print("LACE(CEF)     =", lace_cef)
print("LACE(Riemann) =", lace_riemann)
\end{verbatim}
\end{minipage}\\
\bottomrule
\end{tabular}
\end{table}

\subsection*{Julia: ACE/LMCE/LACE Computation}

\begin{table}[H]
\centering
\begin{tabular}{l}
\toprule
\small
\makecell[l]{\textbf{Example:} Compute carbon metrics using \texttt{PowerModels.jl}}\\
\midrule
\begin{minipage}[t]{0.96\linewidth}\scriptsize
\begin{verbatim}
using PowerModels
using Ipopt

# --- Build carbon-enriched case (Appendix A) ---
net = PowerModels.parse_file("case118.m")
fuel_dict = fuel_dict_generation(net)
carbon_casefile(net, fuel_dict)

# --- Compute carbon metrics ---
R_tot = compute_Rtot(net, result)
ace   = compute_ACE(net, result)

# Locational metrics with selectable backends:
lmce_diff = compute_LMCE(net, result; method="diff")
lmce_mpp  = compute_LMCE(net, result; method="MPP")

lace_cef     = compute_LACE(net, result; method="CEF")
lace_riemann = compute_LACE(net, result; method="Riemann")

println("R_tot = ", R_tot)
println("ACE   = ", ace)
println("LMCE(diff) = ", lmce_diff)
println("LMCE(MPP)  = ", lmce_mpp)
println("LACE(CEF)     = ", lace_cef)
println("LACE(Riemann) = ", lace_riemann)
\end{verbatim}
\end{minipage}\\
\bottomrule
\end{tabular}
\end{table}

\end{document}